\documentclass[a4paper,11pt]{article}
\pdfoutput=1 % if your are submitting a pdflatex (i.e. if you have
\usepackage{jheppub} % for details on the use of the package, please
\usepackage{multirow}
\newcommand{\ba}{\begin{eqnarray}}
\newcommand{\ea}{\end{eqnarray}}

\usepackage[T1]{fontenc} % if needed

\title{\boldmath Cosmological dynamics of interacting dark energy and dark matter in viable models of f(R) gravity}

\author[a]{Daris Samart,}
\author[a]{Bhuddhanubhap Silasan,}
\author[b,c,d]{Phongpichit Channuie}

\affiliation[a]{Khon Kaen Particle Physics and Cosmology Theory Group (KKPaCT), Department of Physics, Faculty of Science, Khon Kaen University, 123 Mitraphap road, Khon Kaen, 40002, Thailand}
\affiliation[b]{College of Graduate Studies, Walailak University, Thasala, Nakhon Si Thammarat, \\80160, Thailand}
\affiliation[c]{School of Science, Walailak University, Thasala, Nakhon Si Thammarat, 80160, Thailand}
\affiliation[d]{Research Group in Applied, Computational and Theoretical Science (ACTS), Walailak University, Thasala, Nakhon Si Thammarat, 80160, Thailand}

\emailAdd{darisa@kku.ac.th,bhuddhanubhaps@kkumail.com,channuie@gmail.com}

\abstract{In this work, we investigate the dynamics of the interacting dark energy and dark matter in viable models of $f(R)$ gravity by using a standard framework of dynamical system analysis. A simple form of the interacting dark energy $Q=3\alpha H\rho_{m}$ is used to study three viable models of $f(R)$ gravity which are consistent with local gravity constraints and satisfying conditions for the cosmological viability. As a result, we find that the fixed points are slightly modified from those obtained in the standard non-interacting analysis of $f(R)$ gravity proposed so far in the literature. In our models of adding this interaction, we find that the dynamical profiles of the universe in the viable $f(R)$ dark energy models are modified by the interaction term as well as their relevant model parameters. Moreover, our results yield the correct cosmological evolution with additional constraint parameter, $\alpha$, from the interacting dark energy.}
\begin{document} 
\maketitle
\flushbottom

%%%%%%%%%%%%%%%%%%%%%%%
\section{Introduction}
%%%%%%%%%%%%%%%%%%%%%%%

The observational evidence that the universe had entered a period of accelerated expansion has been supported by independent observational data such as the Supernovae-type Ia (SN Ia) \cite{Riess,Perlmutter}, the Cosmic Microwave Background (CMB) temperature anisotropies observed by WMAP \cite{WMAP1,WMAP7}, and Baryon Acoustic Oscillations \cite{BAO1,BAO2}. In light of the framework of general relativity (GR), the accelerated expansion is driven by a new energy density component with negative pressure. The unknown component giving rise to this late time cosmic acceleration is termed dark energy (DE). However, little is known about this DE component. The origin of DE responsible for the present time cosmic acceleration is one of unsolved problems in modern cosmology. This phase of cosmic acceleration cannot be explained by the standard equation of state $w=p/\rho$ satisfying the condition $w\geq 0$ with $p$ and $\rho$ being the pressure
and the energy density of matter, respectively. In fact, some unknown component having negative pressure with $w<-1/3$ is needed to describe the acceleration of the universe. The nature of DE is still unknown yet. Many efforts have been made to solve this serious problem, see e.g., Refs.\cite{review1,review2,review3,review4,reviewSahni,review5,Noreview,Woodard,Durrer,Lobo,review6,Braxreview,review7,review8,Nojiri:2010wj}. The simplest candidate for dark energy is the cosmological constant, $\Lambda$. However, as mentioned in Ref.\cite{Weinberg:1988cp}, the cosmological constant is many orders of magnitude smaller than that estimated in modern theories of elementary particles. Indeed, people use dynamical models to distinguish other sources dynamically changing in time from the cosmological constant by considering the evolution of the equation of state of DE. Hence one needs to find some mechanism to obtain a small value of $\Lambda$ to reconcile with observations. 

It is well known that the pure cosmological constant cannot be responsible for the accelerated expansion in the very early universe. This is so since it cannot be connect to the radiation-dominated universe. Yet another compelling candidate responsible for DE as well as inflation is a scalar field with a slowly varying potential. In the context of dark energy, many scalar field models have been constructed. These include scenarios of quintessence \cite{Fujii1982,Ford1987,Wetterich1988,Ratra1988,Caldwell1998,Dvali:2001dd} and k-essence \cite{Chiba2000,Armendariz2000,Tian:2021omz,Odintsov:2020qyw}. They predict a wide variety of
the variation of the equation of state of DE. However, those models cannot be distinguished from the $\Lambda$-cold-dark-matter ($\Lambda$CDM) model using  the current observational data. Additionally, because of a very tiny scalar mass ($m_{\phi}\sim 10^{-33}\,{\rm eV}$) required for the cosmic acceleration today \cite{Carroll1998,Kolda1999}, viable scalar-field models in the framework of particle physics cannot be easily achieved. 

There exists another approach of dynamical dark energy models to explain the acceleration of the universe based on the large-distance modification of gravity. The models belonging to this class are for instance $f(R)$ gravity \cite{Capozziello2002,Capozziello2003,Carroll2004,Nojiri2003} in which solely Ricci scalar is replaced by more general function, i.e., $f(R)$, scalar-tensor theories \cite{Amendola1999,Uzan1999,Chiba1999,Bartolo2000,Perrotta2000,Riazuelo2002}, Galileon gravity \cite{Nicolis2009} and Gauss-Bonnet gravity \cite{Nojiri2005,Nojiri20051}. The dynamical system analysis has been used to qualitatively study in cosmology for several decades \cite{Wainwright-Ellis(1997), Coley:2003mj} and it is shown that this framework is very useful to identify and classify asymptotic behaviors of the cosmological models. Recently the dynamical system is used to investigate the dynamics of various dark energy models (see review \cite{review5} and for $f(R)$ gravity \cite{Bahamonde:2017ize}). The systematic dynamical system analysis in $f(R)$ theories of gravity has been done by a series of papers found in Refs.\cite{Amendola:2006kh,Amendola:2006we,Amendola:2006eh,Tsujikawa:2010zza,Amendola:2007nt,Tsujikawa:2007xu} to identify and classify the appropriate models with a correct cosmological evolution from a huge number of the $f(R)$ gravity models. As the results, the cosmological viable models of $f(R)$ gravity have been constructed with the appropriate trajectories in the dynamical system phase space. In addition, these viable models of $f(R)$ gravity are compatible with the local gravity constraints, see more detail discussions and example in Refs.\cite{review7,review8,Tsujikawa:2010zza}. It is worth noting that an attractive feature of these models is that the cosmic acceleration can be realized without invoking a DE matter component. Moreover, in the light of those models, tight constraints coming from local gravity tests as well as a number of observational constraints are rigorous compared to modified matter models. 

The phase space of $f(R)$ gravity using the autonomous dynamical systems approach was initiated and extended by the authors of Ref.\cite{Odintsov:2017tbc,Oikonomou:2019boy,Chatzarakis:2019fbn,Oikonomou:2017ppp,Odintsov:2019evb}. Although there is a number of work on dynamical system in $f(R)$ gravity with non-interacting dark energy and dark matter, see examples and references therein \cite{Bahamonde:2017ize}, a study of interacting dark energy of $f(R)$ gravity counterpart is less attention \cite{Poplawski:2006kv,Poplawski:2006ey,He:2011qn,He:2015bzs,LHuillier:2017pdi}, in particular for dynamical system analysis \cite{Odintsov:2018uaw}. The study of interacting dark energy and dark matter is useful for understanding accelerated scaling solution behavior between dark energy and dark matter under the conditions $\Omega_{DE}/\Omega_{m}\approx 1,\,\ddot a >0$ with $\Omega_{DE,m}$ being the energy densities of dark energy and dark matter and $\ddot a$ an accelerating rate of the universe's expansion. In addition, the investigation of the interacting dark energy and dark matter might describe a coupling between the structure formation of the dark matter and the time evolution of the dark energy. In other words, it might be helpful to explain why dark energy dominates dark matter at the late time of the universe. Notice that the effect of interacting dark energy models in global 21cm signal was studied in Ref.\cite{Halder:2021jiv}. Some interacting dark energy models and the occurrence of future singularities have been analysed in Ref.\cite{BeltranJimenez:2016dfc}.

The main purpose of this work is to study the interacting dark energy and dark matter in the viable models of $f(R)$ gravity in the following aspects. On the one hand, we do ask the question whether adding the interaction to the viable $f(R)$ models is worth for providing the correct behavior of the cosmological evolution compared to the non-interacting one? On the other hand, we would like to see how the interacting dark energy might modify the parameters in the viable $f(R)$ gravity models.

The structure of the present work is organized as follows. In Sec.\ref{sec2}, we derive cosmological equations and study dynamical system set up in the framework of $f(R)$ gravity. Here we also introduce autonomous system of interacting dark energy and dark matter in $f(R)$ gravity. In section \ref{sec3}, we study fixed points obtained the autonomous system and analyze stability of such fixed points. As a simple model of the interacting dark energy $Q=3\alpha H\rho_{m}$, in Sec.\ref{sec4} we employ it to study three viable models of $f(R)$ gravity which is consistent with local gravity constraints and satisfying conditions for the cosmological viability. In this section, cosmological implications in viable models of $f(R)$ gravity are given. The last section is devoted to conclusions.

%%%%%%%%%%%%%%%%%%%%%%%
\section{Cosmological equations and dynamical system set up}\label{sec2}
%%%%%%%%%%%%%%%%%%%%%%%
\subsection{Background dynamics of the cosmological equations in $f(R)$ gravity}
In this work, we will work on the standard cosmological background with the flat FRW line element, it reads
\begin{eqnarray}
ds^2 = -dt^2+a(t)^2\big(dx^2 + dy^2 + dz^2\big).
\label{FRW-metric}
\end{eqnarray}
While the action of the $f(R)$ gravity and the matter is given by
\begin{eqnarray}
S = \frac{1}{2\kappa^2}\int d^4x \sqrt{-g}\,f(R) + \int d^4x\,\mathcal{L}_{\rm matter},
\label{action}
\end{eqnarray}
where $g$ is determinant of the metric tensor $g_{\mu\nu}$, $\kappa = \sqrt{8\pi G}$ with $G$ is gravitational Newton's constant and $\mathcal{L}_{\rm matter}$ is Lagrangian density of the matter fields which will be considered as the perfect fluid. Varying the action in Eq.(\ref{action}) with respect to the inverse metric tensor $g^{\mu\nu}$, the Einstein field equation of the $f(R)$ gravity is written by,
\begin{eqnarray}
R_{\mu\nu} F -\frac12\,g_{\mu\nu} f -\nabla_\mu\nabla_\nu F +g_{\mu\nu} \Box F = \kappa^2 T_{\mu\nu}\,,
\label{EFE}
\end{eqnarray}
where $F \equiv f_R = \partial f/\partial R$ and $\Box \equiv \nabla_\mu\nabla^\mu$\. The energy momentum tensor $T_{\mu\nu}$ is defined by
\begin{eqnarray}
T_{\mu\nu} = -\frac{2}{\sqrt{-g}}\frac{\delta \mathcal{L}_M}{\delta g^{\mu\nu}}= \big(\rho_m + \rho_r \big)u_\mu u_\nu - p_r g_{\mu\nu}\,
\end{eqnarray}
where $u^\mu=\big(1,0,0,0\big)$ is four velocity in the co-moving frame. $\rho_{m,r}$ are energy density of the dark (and dust) matter and radiation respectively, $p_r$ is the radiation pressure that obeys the equation of state (EoS) as $p_r=\rho_r/3$\,. Here we included the dark matter and dust as the same matter specie for simplicity. According to field equation in (\ref{EFE}) with the FRW metric in (\ref{FRW-metric}), one finds
\begin{eqnarray}
3FH^2 &=& \frac12(FH -f) -3 H \dot F + \kappa^2(\rho_m + \rho_r)\,,
\label{Friedmann}
\\
2F\dot H &=& H\dot F - \ddot F -\kappa^2\Big( \rho_m + \frac43 \rho_r\Big)\,,
\label{raychaudhuri}
\end{eqnarray}
where $H\equiv \dot a/a$ is the Hubble parameter and Ricci scalar from FRW metric is $R = 6\big( 2H^2 + \dot H\big)$ while $\dot~ \equiv d/dt$ is represented the cosmic time derivative. More importantly, one can rewrite the Friedmann and Raychuadhuri equations in the following forms \cite{Amendola:2006we}, 
\begin{eqnarray}
3AH^2 &=& \kappa^2\big( \rho_{DE} + \rho_m +\rho_r\big)\,,
\\
2A\dot H &=& -\kappa^2\Big( \rho_{DE} + p_{DE} + \rho_m +\frac43\,\rho_r \Big)\,,
\end{eqnarray}
where $A$ is arbitrary real constant and $\rho_{DE}$ and $p_{DE}$ are defined by
\begin{eqnarray}
\rho_{DE} &\equiv& \frac12(FR-f)-3H\dot F + 3H^2(A-F)\,, 
\\
p_{DE} &\equiv&  \ddot F + 2H\dot F -\frac12\,(FR-f)-(3H^2+2\dot H)(A-F)\,,
\end{eqnarray}
see more detail discussions of physical meaning of $\rho_{DE}$ and $p_{DE}$ in \cite{Amendola:2006we,Tsujikawa:2010zza,review7}. We might call the quantities $\rho_{DE}$ and $p_{DE}$ as effective energy density and pressure of the curvature fluid. It was shown in Refs. \cite{Amendola:2006we,review7} that $\rho_{DE}$ and $p_{DE}$ obey the conservation equation as
\begin{eqnarray}
\dot\rho_R + 3 H\big(\rho_{DE} + p_{DE}\big) &=& 0
\label{conserved-R}
\end{eqnarray}
By considering the conservation of energy momentum tensor, $\nabla^\mu\, T_{\mu\nu} =0$\,, we find
\begin{eqnarray}
\dot\rho_m + 3 H\rho_m &=& 0\,,
\label{conserved-matter}
\\
\dot\rho_r + 4 H\rho_r &=& 0\,.
\label{conserved-radiat}
\end{eqnarray}
We observe that the evolution equations in Eqs. (\ref{conserved-R},\ref{conserved-matter},\ref{conserved-radiat}) are conserved separately. To study the interaction between dark energy and dark matter, however, we can modify the conservation equations of the dark energy of the curvature fluid and the matter in Eqs. (\ref{conserved-R}) and (\ref{conserved-matter}) by including some coupling term, $Q$ as
\begin{eqnarray}
\dot\rho_R + 3 H\big(\rho_{DE} + p_{DE}\big) &=& Q\,,
\label{interact-R}
\\
\dot\rho_m + 3 H\rho_m &=& -Q\,.
\label{interact-m}
\end{eqnarray}
Although the above equations do not follow the conservation law separately but the the total energy density $\rho_{\rm total} = \rho_{DE} + \rho_m + \rho_r$ does conserve without violating the total energy momentum tensor. In addition, the coupling dark energy and dark matter term $Q$ can be interpreted as the exchange rate of the energy density between dark energy and dark matter. The signs of $Q$ also reflect that $Q>0$ means the energy density of dark matter transfer to dark energy whereas $Q<0$ the energy density of dark energy will be transferred to dark matter. It is worth to note that the physical range of the $\alpha$ parameter should be $\alpha<-1/3$ in order to prevent the dark (and dust) matter decaying slower than the radiation.    

In this work, we will examine the interacting dark energy in viable $f(R)$ gravity models by using a simple coupling term as 
\begin{eqnarray}
Q = 3\alpha H\rho_m\,.
\label{interacting-Q}
\end{eqnarray}
This model is propose in Ref.\cite{Chimento:2003iea,Boehmer:2008av,Chen:2008pz} and mostly used to study cosmological evolution of interacting dark energy and dark matter in the dynamical system analysis. We have completed deriving all important cosmological equation that are crucial for dynamical system analysis. In the next section, we will set up the autonomous equations of the dynamical system of the interacting dark energy and dark matter.

\subsection{Autonomous system of interacting dark energy and dark matter in $f(R)$ gravity}
In this section, we will derive the autonomous equations which are the main ingredients for studying the dynamical system in the interacting dark energy and dark matter in general $f(R)$ gravity. After setting up the dynamical equations, we then find the fixed points of the system. 

According to the Friedmann equation in Eq. (\ref{Friedmann}), we can define the dimensionless variable as \cite{Amendola:2006we},
\begin{eqnarray}
x_1=-\frac{\dot{F}}{HF}\,,\quad x_2 = -\frac{f}{6FH^2}\,,\quad
x_3=\frac{R}{6H^2}=\frac{\dot{H}}{H^2}+2\,,\quad x_4=\frac{\kappa^2\rho_r}{3FH^2}\,.
\label{dimensionless}
\end{eqnarray}
The autonomous system is given by
\begin{eqnarray}
\frac{d x_1}{d N} &=&-1-x_3-3x_2+x_1^2-x_1x_3+x_4 -9\alpha(1-x_1-x_2-x_3-x_4)\,,
\nonumber\\
\frac{d x_2}{d N} &=&\frac{x_1x_3}{m}-x_2(2x_3-4-x_1)\,,
\nonumber\\
\frac{d x_3}{d N} &=&-\frac{x_1x_3}{m}-2x_3(x_3-2)\,,
\nonumber\\
\frac{d x_4}{d N} &=&-2x_3x_4+x_1x_4\,,
\label{autonomous}
\end{eqnarray}
where the variable $N\equiv \ln a$ is e-folding number and leads to $d/dN\equiv d/(Hdt)$\,. The parameter $m$ is defined by
\begin{eqnarray}
m = m(r) \equiv \frac{d\ln F}{d\ln R}=\frac{R F_R}{F}\,,\qquad\quad
r \equiv -\frac{d\ln f}{d\ln R} = -\frac{R F}{f}= \frac{x_3}{x_2}\,
\label{define-m-r}
\end{eqnarray}
with $F_R=dF/dR$ and it has been explained in Ref.\cite{Amendola:2006we} that the parameter $m$ is represented the deviation of the $f(R)$ gravity from the LCDM. For example, $f(R)= R -2\Lambda$ leads to $m=0$\,. In addition, the density parameters for each species of matters can be related to the dimensionless variables in Eq. (\ref{dimensionless}) by the Friedmann equation in Eq. (\ref{Friedmann}) as the following constraint equations, 
\begin{eqnarray}
\Omega_m \equiv \frac{\kappa^2\rho_m}{3FH^2} = 1 - x_1 -x_2 - x_3 - x_4\,,
\qquad
\Omega_r \equiv x_4\,,\qquad \Omega_{DE} \equiv x_1 + x_2 + x_3\,.
\label{densitypara}
\end{eqnarray}
Finally, the effective EoS is written in terms of the dimensionless variable by
\begin{eqnarray}
w_{\rm eff} = -1-\frac{2\dot H}{3 H^2}=-\frac13(2x_3-1)\,.
\end{eqnarray}
The effective EoS parameter is also used to identify that the universe is accelerated expansion when $w_{\rm eff}<-1/3$\,. We will use the derived autonomous system in this section to determine the fixed points and evaluate its eigenvalues for analyzing the dynamical system of the interacting dark energy and dark matter in the next section.

%%%%%%%%%%%%%%%%%%%%%%%
\section{Fixed points and stability analysis}\label{sec3}
%%%%%%%%%%%%%%%%%%%%%%%
In the previous section, we have already set up the autonomous system and we are now ready to determine the fixed points from the autonomous system in Eq. (\ref{autonomous}) directly by setting $dx_i/dN = 0$ with $i=1,\,2,\,3,\,4$\,. We find that there are total eight fixed points and they are classified into two cases as absent of the radiation and non-vanished radiation in the universe as found in Ref.\cite{Amendola:2006we} for the dynamical system analysis of the non-interacting dark energy in $f(R)$ gravity. In addition, all fixed points found in this work are modified by the interacting dark energy parameter $\alpha$ and they will be reduced down to all fixed points in Ref.\cite{Amendola:2006we} when $\alpha\to 0$. Therefore, we will call all fixed points following Ref.\cite{Amendola:2006we} with slight distinction. The fixed points and its eigenvalues of the stability matrix will be determined in the following subsections. 
%%%%%%%%%%%%%%%%%%%%%%%%%%%%%%%%%%%%%%%%%%%%%%%%%%%%%%%%%%%%%%%%%%%%%%%%%%%%%%%%%%%%%%%%%%%%%%%%%%%%%%%%%%%%%%%%%%%%%%%%%%%%%%%%%%%%%%%%%%%%%%%%%%%%%%%%%%
\subsection{$P_1$: the de Sitter fixed point} 
\begin{align}
P_1:(x_1,x_2,x_3,x_4)&=(0,-1,2,0)
\end{align}
The dark matter density and EoS parameters are given by
\begin{align}
\Omega_\text{m}=0,\qquad w_\text{eff}=-1.
\end{align}
In this point, both dark matter and radiation are absent while the universe is dominated by dark energy. EoS of this fixed point indicates that the universe is in the accelerating expansion phase. To study the stability of the fixed point, we can use the eigenvalues of the stability matrix to characterize the behaviors of the fixed point. The stability matrix is defined by 
\begin{eqnarray}
\mathcal{M} = \left(
\begin{array}{c c c c}
\frac{\partial Y_1}{\partial x_1} & \frac{\partial Y_1}{\partial x_2} & \frac{\partial Y_1}{\partial x_3} & \frac{\partial Y_1}{\partial x_4}
\\
\frac{\partial Y_2}{\partial x_1} & \frac{\partial Y_2}{\partial x_2} & \frac{\partial Y_2}{\partial x_3} & \frac{\partial Y_2}{\partial x_4}
\\
\frac{\partial Y_3}{\partial x_1} & \frac{\partial Y_3}{\partial x_2} & \frac{\partial Y_3}{\partial x_3} & \frac{\partial Y_3}{\partial x_4}
\\
\frac{\partial Y_4}{\partial x_1} & \frac{\partial Y_4}{\partial x_2} & \frac{\partial Y_3}{\partial x_3} & \frac{\partial Y_4}{\partial x_4}
\end{array}
\right),
\end{eqnarray}
where $Y_i\equiv dx_i/dN$\,. The eigenvalues for fixed point $P_1$ read
\begin{align}
-\frac{3}{2}\pm\frac{\sqrt{25-16/m}}{2}\,,~3(3\alpha-1)\,,~0\,.
\end{align}
The stability behaviors of this fixed point are as follows: stable node or attractor fixed point for $0<m\leq 1$ and $\alpha <1/3$. Saddle point for otherwise.
%%%%%%%%%%%%%%%%%%%%%%%%%%%%%%%%%%%%%%%%%%%%%%%%%%%%%%%%%%%%%%%%%%%%%%%%%%%%%%%%%%%%%%%%%%%%%%%%%%%%%%%%%%%%%%%%%%%%%%%%%%%%%%%%%%%%%%%%%%%%%%%%%%%%%%%%%%
\subsection{$P_2$: Modified $\phi$-Matter Dominate Epoch (M$\phi$-DME) point }
This fixed point is similar to the fixed point in non-interacting dark energy in $f(R)$ gravity case found in Ref.\cite{Amendola:2006we}. But the fixed point $P_2$ is slightly modified by the interacting dark energy parameter, $\alpha$. It reads
\begin{align}
P_2:(x_1,x_2,x_3,x_4)&=(-1-9\alpha,0,0,0).
\end{align}
Matter density and EoS parameters are given by
\begin{align}
\Omega_\text{m}=2+9\alpha,\qquad w_\text{eff}=\frac{1}{3}.
\end{align}
The eigenvalues of the stability matrix are read
\begin{align}
-3 (3 \alpha -1)\,,~-9 \alpha -2,\frac{1 + 9 \alpha +4 m}{m}\,,~0\,.
\end{align}
The conditions of stability are as follows : stable node for $(1/4)(-1-9\alpha)<m<0$ and $\alpha>1/3$. Unstable node for $m<0$ or $m>(1/4)(-1-9\alpha)$ and $\alpha<-2/9$. We note that the fixed point $P_2$ is the stable node when $\alpha>1/3$. This contradicts to the physical range of the interacting dark energy parameter ($\alpha$), i.e. $\alpha<1/3$, which makes dark matter decaying faster than the radiation. 
%%%%%%%%%%%%%%%%%%%%%%%%%%%%%%%%%%%%%%%%%%%%%%%%%%%%%%%%%%%%%%%%%%%%%%%%%%%%%%%%%%%%%%%%%%%%%%%%%%%%%%%%%%%%%%%%%%%%%%%%%%%%%%%%%%%%%%%%%%%%%%%%%%%%%%%%%%
\subsection{$P_3$: Pure kinetic dominated point}
The fixed point $P_3$ has the same form as the non-interacting $f(R)$ gravity case \cite{Amendola:2006we}. It reads
\begin{align}
P_3:(x_1,x_2,x_3,x_4)&=(1,0,0,0).
\end{align}
The dark matter density and EoS parameters are read
\begin{align}
\Omega_\text{m}=0,\qquad w_\text{eff}=\frac{1}{3}.
\end{align}
The fixed point $P_3$ shows that the universe is dominated by dark energy as the $P_1$ point with the decelerating universe due to $w_\text{eff}=1/3$ and making $\ddot a<0$\,. The eigenvalues of the fixed point $P_3$ are given by
\begin{align}
5\,,~4-\frac{1}{m}\,,~9 \alpha +2 \,.
\end{align}
The conditions of stability are as follows : unstable node for $m<0$ or $m>1/4$ and $\alpha>-2/9$, otherwise is for saddle point.
%%%%%%%%%%%%%%%%%%%%%%%%%%%%%%%%%%%%%%%%%%%%%%%%%%%%%%%%%%%%%%%%%%%%%%%%%%%%%%%%%%%%%%%%%%%%%%%%%%%%%%%%%%%%%%%%%%%%%%%%%%%%%%%%%%%%%%%%%%%%%%%%%%%%%%%%%%
\subsection{$P_4$: Additional pure kinetic dominated point}
The fixed point $P_4$ appears the same as the non-interacting dark energy $f(R)$ gravity case \cite{Amendola:2006we}. It is given by
\begin{align}
P_4:(x_1,x_2,x_3,x_4)&=(-4,5,0,0).
\end{align}
The dark matter density and the EoS parameters are 
\begin{align}
\Omega_\text{m}=0,\qquad w_\text{eff}=\frac{1}{3}.
\end{align}
The fixed point $P_4$ has the same properties as the $P_3$ point while eigenvalues of the $P_4$ point are given by
\begin{align}
-5\,,~\frac{4 (m+1)}{m}\,,~3(3\alpha-1)\,,~0\,.
\end{align}
The conditions of stability are as follows : stable node for $-1<m<0$ and $\alpha<1/3$, unstable node for $m<-1$ or $m>0$ and $\alpha>1/3$, otherwise is for saddle point.
%%%%%%%%%%%%%%%%%%%%%%%%%%%%%%%%%%%%%%%%%%%%%%%%%%%%%%%%%%%%%%%%%%%%%%%%%%%%%%%%%%%%%%%%%%%%%%%%%%%%%%%%%%%%%%%%%%%%%%%%%%%%%%%%%%%%%%%%%%%%%%%%%%%%%%%%%%
\subsection{$P_5$: scaling solution point}
The fixed point $P_5$ represents scaling solution point of the universe. The scaling solution gives to the ratio $\Omega_m/\Omega_{DE}$ becoming constant. It has been pointed out in Ref.\cite{Amendola:2006we} that the fixed point $P_5$ represents the standard matter when $m\to 0$\,. In addition, it is worth noting that our result modifies the scaling solution of $f(R)$ gravity with the interacting dark energy parameter $\alpha$\,. The fixed point $P_5$ reads
\begin{align}
P_5:(x_1,x_2,x_3,x_4)&=\left(\frac{(3-9 \alpha ) m}{m+1},-\frac{9 \alpha +4 m+1}{2 (m+1)^2},\frac{9 \alpha +4 m+1}{2 m+2},0\right).
\end{align}
The dark matter density and the EoS parameters are
\begin{align}
\Omega_\text{m}=\frac{2 (9 \alpha -4) m^2+(9 \alpha -3) m+2}{2 (m+1)^2},\qquad w_\text{eff}=-\frac{3 \alpha +m}{m+1},
\end{align}
where the dark matter density is greater than zero under the following condition
\begin{align}
-\frac{1}{4} \sqrt{\frac{81 \alpha ^2-198 \alpha +73}{(9 \alpha -4)^2}}-\frac{3 (3 \alpha -1)}{4 (9 \alpha -4)}\leq m\leq \frac{7-9 \alpha }{18 \alpha -10}.
\end{align}
The condition for accelerating universe is given by
\begin{align}
m<-1\qquad\text{or}\qquad m>\frac{1}{2} (1-9 \alpha ).
\end{align}
The eigenvalues of the $P_5$ read
\begin{align}
\frac{-18 \alpha  m^{3/2}-3 (3 \alpha +1) \sqrt{m} \pm \sqrt{-16 (9 \alpha +1)+4 p m^3+ q m^2+ s m}}{4 \sqrt{m} (m+1)}\,,~3(1-3 \alpha)\, ,~0\,,
\end{align}
where $s=(8-9 \alpha )^2$, $q=\left(-972 \alpha ^2+252 \alpha +160\right)$ and $s= \left(-567 \alpha ^2+198 \alpha -31\right)$.
There are no ranges of $\alpha$ and $m$ (assuming both are real number) of the eigenvalues of the $P_5$ fixed point for stable and unstable nodes. Then $P_5$ point is always saddle point under the following ranges of the parameters,
\begin{eqnarray}
&&\alpha \leq -\frac{2}{9}\land \left(0<m<\frac{1}{4} \sqrt{\frac{81 \alpha ^2-198 \alpha +73}{(9 \alpha -4)^2}}-\frac{3 (3 \alpha -1)}{4 (9 \alpha -4)}\lor m>-\frac{1}{4} (9 \alpha +1)\right),
\nonumber\\
&&-\frac{2}{9}<\alpha \leq -\frac{1}{9}\land \left(0<m<-\frac{1}{4} (9 \alpha +1)\lor m>\frac{1}{4} \sqrt{\frac{81 \alpha ^2-198 \alpha +73}{(9 \alpha -4)^2}}-\frac{3 (3 \alpha -1)}{4 (9 \alpha -4)}\right),
\nonumber\\
&&-\frac{1}{9}<\alpha <\frac{1}{3}\land m>\frac{1}{4} \sqrt{\frac{81 \alpha ^2-198 \alpha +73}{(9 \alpha -4)^2}}-\frac{3 (3 \alpha -1)}{4 (9 \alpha -4)}\,.
\end{eqnarray}
We note that the energy transfer parameter, $\alpha$ can be both positive and negative. In the other word, the dark matter transfer energy density to dark energy (positive $\alpha$) and dark energy can also transfer energy density to dark matter (negative $\alpha$) in the scaling solution, $P_5$ point. 
%%%%%%%%%%%%%%%%%%%%%%%%%%%%%%%%%%%%%%%%%%%%%%%%%%%%%%%%%%%%%%%%%%%%%%%%%%%%%%%%%%%%%%%%%%%%%%%%%%%%%%%%%%%%%%%%%%%%%%%%%%%%%%%%%%%%%%%%%%%%%%%%%%%%%%%%%%
\subsection{$P_6$: Curvature dominated point}
The fixed point $P_6$ can be used to explain the late accelerating expansion of the universe by the curvature effect of the $f(R)$ gravity. The $P_6$ point is given by
\begin{align}
P_6:(x_1,x_2,x_3,x_4)&=\left(\frac{2-2 m}{2 m+1},\frac{1-4 m}{m (2 m+1)},\frac{4 m^2+3 m-1}{m (2 m+1)},0\right).
\end{align}
The dark matter and EoS parameters of the $P_6$ point are 
\begin{align}
\Omega_\text{m}=0,\qquad w_\text{eff}=\frac{-6 m^2-5 m+2}{6 m^2+3 m}.
\end{align}
The accelerating universe condition $(w_{\rm eff}<-1/3)$ is constrained by 
\begin{align}
m<-\frac{1}{2} \left(\sqrt{3}+1\right)\quad\text{or}\quad -\frac{1}{2}<m<0\quad\text{or}\quad m>\frac{1}{2} \left(\sqrt{3}-1\right).
\end{align}
In addition, the curvature dominated point $P_6$ is reduced to the de Sitter fixed point with $w_{\rm eff}=-1$ at the limit $m\to 0$\,. Universe is entirely dominated by dark energy without any matter. The eigenvalues of the stability matrix for $P_6$ point are given by
\begin{align}
-4+\frac{1}{m}\,,~\frac{2(1- m^2)}{m(2 m +1)}\,,~\frac{18 \alpha  m^2-8 m^2+9 \alpha  m-3 m+2}{m(2 m+1)}\,,~0\,.
\end{align}
The conditions of stability are as follows : stable node for $m<-1$ or $-1/2<m<0$ or $m>1$ and $\alpha<1/3$, while saddle point is otherwise.
%%%%%%%%%%%%%%%%%%%%%%%%%%%%%%%%%%%%%%%%%%%%%%%%%%%%%%%%%%%%%%%%%%%%%%%%%%%%%%%%%%%%%%%%%%%%%%%%%%%%%%%%%%%%%%%%%%%%%%%%%%%%%%%%%%%%%%%%%%%%%%%%%%%%%%%%%%
\subsection{$P_7$: Standard radiation point}
The fixed point $P_7$ has the same values as the non-interacting $f(R)$ gravity case. It characterizes the standard radiation dominated era in the universe. The fixed point $P_7$ is given by
\begin{align}
P_7:(x_1,x_2,x_3,x_4)&=(0,0,0,1).
\end{align}
The dark matter density and EoS parameters are given by
\begin{align}
\Omega_\text{m}=0,\qquad w_\text{eff}=\frac{1}{3}.
\end{align}
In addition the $P_7$ point gives $\Omega_r = 1$\,. The eigenvalues of the $P_7$ are 
\begin{align}
4\,,~4\,,~-1\,,~1+ 9 \alpha\,.
\end{align}
According to the $P_7$ point's eigenvalues, this means the $P_7$ point is saddle point obviously.
%%%%%%%%%%%%%%%%%%%%%%%%%%%%%%%%%%%%%%%%%%%%%%%%%%%%%%%%%%%%%%%%%%%%%%%%%%%%%%%%%%%%%%%%%%%%%%%%%%%%%%%%%%%%%%%%%%%%%%%%%%%%%%%%%%%%%%%%%%%%%%%%%%%%%%
\subsection{$P_8$: $\phi$-radiation Dominate Epoch ($\phi$-RDE) point}
The fixed point $P_8$ takes the same form as the non-interacting $f(R)$ gravity because we do not couple the interaction of the dark energy to the radiation. In addition, this fixed point contains non-vanished dark energy component in the universe. The fixed point $P_8$ reads
\begin{align}
P_8:(x_1,x_2,x_3,x_4)&=\left(\frac{4m}{1+m},-\frac{2m}{(1+m)^2},\frac{2m}{1+m},\frac{1-2m-5m^2}{(1+m)^2}\right).
\end{align}
The dark matter density and EoS parameters are given by
\begin{align}
\Omega_\text{m}=0,\qquad w_\text{eff}=\frac{1-3m}{3+3m}.
\end{align}
According to the EoS, the $P_8$ is reduced to the standard radiation at the limit $m\to 0$\,. The eigenvalues of the $P_8$ are given by
\begin{align}
4\,,~\frac{m^2-1 \pm (m+1) \sqrt{3\big(27 m^2+10 m-5\big)}}{2 (m+1)^2}\,,~1+9 \alpha\,.
\end{align}
This shows clearly that the $P_8$ point is always saddle point.

As a result, we note that there are two fixed points $P_1$ and $P_6$ that can be represented as the late time accelerating universe in $f(R)$ gravity; while the $P_5$ is only one fixed point for the (dark) matter dominated epoch and then transits to late time dark energy dominated era. It has been shown in Ref.\cite{Amendola:2006we} that there are two classes of cosmological viability in $f(R)$ gravity models, i.e., class A saddle point $P_5$ transits to stable node $P_1$ de Sitter fixed point and class B saddle point $P_5$ transits to the curvature dominated $P_6$ point for non-interacting $f(R)$ gravity. However, the parameter $\alpha$ of the interacting term of the $Q=3\alpha H \rho_m$ will play the important roles on the cosmological viability and the estimations of the transition from the matter dominated era (saddle point) to the dark energy domination (stable node) at the late time of the viable $f(R)$ gravity models. In the next section, we will examine the cosmological dynamics for each viable $f(R)$ gravity models in detail.
%%%%%%%%%%%%%%%%%%%%%%%%%%%%%%%%%%%%%%%%%%%%%%%%%%%%%%%%%%%%%%%%%%%%%%%%%%%%%%%%%%%%%%%%%%%%%%%%%%%%%%%%%%%%%%%%%%%%%%%%%%%%%%%%%%%%%%%%%%%%%%%%%%%%
\section{Stability and cosmological implications in viable models of $f(R)$ gravity}\label{sec4}
%%%%%%%%%%%%%%%%%%%%%%%%%%
In this section, we will investigate the cosmological dynamics of the interacting dark energy in $f(R)$ gravity for specific viable models of $f(R)$. The viable models of $f(R)$ gravity require a correct cosmological phase of evolution as of the standard LCDM \cite{Amendola:2006kh,Amendola:2006eh}, free of matter \cite{Dolgov:2003px,Faraoni:2006sy} and cosmological \cite{Carroll:2006jn,Song:2006ej,Sawicki:2007tf} instabilities as well as satisfy to the local gravity constraints \cite{ Olmo:2005zr,Chiba:2006jp,Capozziello:2007eu,Olmo:2005hc,Navarro:2006mw}.  First of all, we would like to recall the conditions for the cosmologically viable $f(R)$ dark energy models as given in Refs.\cite{review7,review8}
\begin{eqnarray}
F >0\,,\qquad\quad F_R >0\,, \qquad\quad {\rm for} \quad R\geq R_0 \geq 0\,, 
\end{eqnarray}
where $F_R\equiv \partial F/\partial R = \partial^2 f/\partial R^2$ and $R_0$ is the current value of the Ricci scalar. The condition $F>0$ is required to avoid the appearance of the ghost in the $f(R)$ gravity while the condition $F_R>0$ is to prevent the negative mass of the scalaron field in the Einstein frame. In addition, Ref.\cite{Amendola:2006we} has systematically studied and classified the categories of the classes of $f(R)$ gravity models in the dynamical system approach. We refer readers to Refs.\cite{review7,review8} for excellent review about viable models of $f(R)$ gravity and reference therein. In the following subsection, we will use three cosmologically viable $f(R)$ models which have been demonstrated in \cite{review7,review8} that with the proper ranges of the parameters, the models have a correct cosmological evolution i.e. radiation (saddle point) to dark and dust matter (saddle point) to late time dark energy dominated (stable node). Moreover the following viable models of $f(R)$ gravity also agrees with the local gravity constraints.  
%%%%%%%%%%%%%%%%%%%%%%%%%%%%%%%%%%%%%%%%%%%%%%%%%%%%%%%%%%%%%%%%%%%%%%%%%%%%%%%%%%%%%%%%%%%%%%%%%%%%%%%%%%%%%%%%%%%%%%%%%%%%%%%%%%%%%%%%%%%%%%%%%%%%%%
\subsection{Model A: $f(R) = R - \gamma\,R^n$\,with $\gamma>0$ and $0<n<1$}
We start with the model A: $f(R) = R - \gamma\,R^n$\,with the conditions $\gamma>0$ and $0<n<1$ which provide a correct cosmological evolution. This model was presented in Refs. \cite{Amendola:2006we,Li:2007xn} Firstly, we determine values of $m$ and $r$ from their definitions in Eq.(\ref{define-m-r}). One finds,
\begin{align}
m&=\frac{\gamma  (n-1) n R^n}{R- \gamma  n R^n},\\
r&=-\frac{R -\gamma  n R^n}{R -\gamma  R^n}.
\end{align}
Then we can write $m(r)$ as a function of $r$ as
\begin{align}
m(r)=\frac{n(1+r)}{r} \label{eq:m(r)ofAmodel}
\end{align}
Notice that equation (\ref{eq:m(r)ofAmodel}) does not depend on $\gamma$. Next we determine the values of all fixed point, $\Omega_m$, $w_{\rm eff}$ and their eigenvalues by using the parameter $m$ summarized in table of each fixed point and obtain the set of $r$ given in table \ref{table:modelA}. We find
\allowdisplaybreaks
\begin{eqnarray}
&\bullet& P_1 = (0,-1,2,0)\,,
\nonumber\\
&& \Omega_\text{m}=0,\qquad w_\text{eff}=-1.
\nonumber\\
&&{\rm eigenvalues}~:~-\frac{3}{2}\pm\frac{\sqrt{25-32/n}}{2}\,,~3(3\alpha-1)\,,~0\,.
\\
&\bullet& P_2 = (-1-9\alpha,0,0,0)\,,
\nonumber\\
&&\Omega_\text{m} = 2+9\alpha,\qquad w_\text{eff}=\frac{1}{3}\,,
\nonumber\\
&&{\rm eigenvalues}~:~-3 (3 \alpha -1)\,,~-9 \alpha -2\,,~\frac{1 + 9 \alpha +4 (n-1)}{n-1}\,,~0\,,
\\
&\bullet& P_3 = (1,0,0,0).
\nonumber\\
&&\Omega_\text{m} = 0,\qquad w_\text{eff}=\frac{1}{3}
\nonumber\\
&&{\rm eigenvalues}\quad\,:~5\,,~4-\frac{1}{n-1}\,,~9 \alpha +2 \,.
\\
&\bullet& P_4 = (-4,5,0,0)\,,
\nonumber\\
&&\Omega_\text{m} = 0,\qquad w_\text{eff}=\frac{1}{3}
\nonumber\\
&&{\rm eigenvalues}\quad\,:~-5\,,~\frac{4 (n)}{n-1}\,,~3(3\alpha-1)\,,~0\,.
\\
&\bullet& P_5 = \left(\frac{(3-9 \alpha ) (n-1)}{n},-\frac{9 \alpha +4(n-1)+1}{2 n^2},\frac{9 \alpha +4(n-1)+1}{2n},0\right)
\nonumber\\
&& \Omega_\text{m} = \frac{2 (9 \alpha -4) (n-1)^2+(9 \alpha -3) (n-1)+2}{2 n^2},\qquad w_\text{eff}=-\frac{3 \alpha +n -1}{n},
\nonumber\\
&&{\rm eigenvalues}\quad\,:~\frac{1}{4 \sqrt{n-1} (n)}\Big( -18 \alpha  (n-1)^{3/2}-3 (3 \alpha +1) \sqrt{n-1}
\nonumber\\
&&\qquad\qquad\qquad\qquad ~\pm \sqrt{-16 (9 \alpha +1)+4 p (n-1)^3+ q (n-1)^2+ s (n-1)}\Big)\,,
\nonumber\\
&&\qquad\qquad\qquad\qquad   ~3(1-3 \alpha)\, ,~0\,,
\\
&\bullet& P_6 = \left(-\frac{2 (n-2)}{2 n-1}\,,\,\frac{5-4 n}{2 n^2-3 n+1}\,,\,\frac{n (4 n-5)}{(n-1) (2 n-1)}\,,\, 0\right),
\nonumber\\
&&\Omega_\text{m} = 0,\qquad w_\text{eff}=\frac{-6 m^2-5 m+2}{6 m^2+3 m}
\nonumber\\
&&{\rm eigenvalues}\quad\,:~-4+\frac{1}{n-1}\,,-\frac{2 (n-2) n}{2 n^2-3 n+1}\,,
\nonumber\\
&&\qquad\qquad\qquad\qquad   ~\frac{9 \alpha +2 (9 \alpha -4) n^2+(13-27 \alpha ) n-3}{2 n^2-3 n+1}\,,~0\,,
\\
&\bullet& P_7 = (0,0,0,1)
\nonumber\\
&&\Omega_\text{m} = 0,\qquad w_\text{eff}=\frac{1}{3}
\nonumber\\
&&{\rm eigenvalues}\quad\,:~4\,,~4\,,~-1\,,~1+ 9 \alpha\,.
\\
&\bullet& P_8 = \left(\frac{4(n-1)}{n},-\frac{2(n-1)}{n^2},\frac{2(n-1)}{n},\frac{1-2(n-1)-5(n-1)^2}{n^2}\right)
\nonumber\\
&& \Omega_\text{m} = 0,\qquad w_\text{eff}=\frac{4+3n}{3n}
\nonumber\\
&&{\rm eigenvalues}\quad\,:~4\,,~\frac{n^2 -2 n\pm n \sqrt{3\big(27 n^2-44 n+12\big)}}{2 n^2}\,,~1+9 \alpha\,.
\end{eqnarray}
We have summarized the list of parameters $m$ and $r$ for all fixed points of the model A as well as their stability in table \ref{table:modelA}.

Interestingly, the points $P_{2,3}$ with $r \equiv x_3/x_2 =0/0$ are undefined but one might use the L'Hospital rule to represent the parameter $r$ for $P_{2,3}$ as,
\begin{align}
r\equiv - \frac{R F}{f} = -\frac{\partial (RF)/\partial R}{\partial f/\partial R}= -\frac{R F_R - F}{F}=-m-1\,.
\end{align}
This trick also can be applied to determine the parameter $m$ at the $P_4$ point.

As a result given in table \ref{table:modelA}, the de Sitter fixed point $P_1$ does violate the condition $0<n<1$ for the $f(R)= R -\gamma R^n$ model as shown in \cite{Amendola:2006we} when it becomes the stable node. Only physical stable node in this model is the curvature fluid dominated point $P_6$ with $1/2<n<1$. More importantly, the scaling solution fixed point $P_5$ is not compatible with the condition $0<n<1$ for a saddle point. The range of $n$ lies outside $0<n<1$ for all possible $\alpha$ values, see table \ref{table:modelA}. We find $C_1\approx 0.32,\,3.99,\,4.89$ and $C_2\approx 7.67,\,5.16,\,5.00$ for $\alpha\approx 1/3,\,-1/9,\,-2/9$ respectively. Then we observe that the interacting dark energy term $Q=3\alpha H \rho_m$ does change the profile of the model A and does not provide correct cosmological evolution. In addition, we would like to stress that the model A will be a viable $f(R)$ gravity model when $\alpha \to 0$ recovering the non-interacting DE case as shown in Refs. \cite{review7,Amendola:2006we,Tsujikawa:2010zza}. The model A is apparently sensitive to the inclusion of the interacting DE model $Q=3\alpha H \rho_m$ for any $\alpha \neq 0$. However, other choices of the interacting DE coupling, $Q$, might be worth for further investigation to make the model A viable.
%%%%%%%%%%%%%%%%%%%%%%%%%%%%%%%%%%%%%%%%%%%%%%%%%%%%%%%%%%%%%%%%%%%%%%%%%%%%%%%%%%%%%%%%%%%%%%%%%%%%%%%%%%%%%%%%%%%%%%%%%%%%%%%%%%%%%%%%%%%%%%%%%%%%%%%%%
\renewcommand{\arraystretch}{1.8}
\begin{table}
\begin{center}
\begin{tabular}{ |c|c|c|c|c|c| }
\hline
\multirow{2}{*}{}&	\multirow{2}{*}{$r=x_3/x_2$}	&\multirow{2}{*}{$m(r)$}	&	\multicolumn{3}{ |c | }{Stability} \\
\cline{4-6}
&	&	& Stable & Unstable & Saddle \\ \hline
$P_1$	&	$-2$		&	$n/2$	& 	$\frac{32}{25}<n<2$	&  $-$ & otherwise 
\\
		&			&			&	  $\wedge$ $\alpha\le \frac13$  	&	 &	\\
$P_2$	&	$-1-m$		&	$n-1$	& $\frac{C_1}{4}<n<1$ &  $n<\frac{C_1}{4}$ $\vee$ & otherwise \\
		&				&			& $\wedge$ $\alpha>\frac13$ &  $\Big[n<1$ $\wedge$ $\alpha<-\frac29\Big]$  & \\
$P_3$	&	$-1-m$		&	$n-1$	& $n<1$ $\vee$ $n>\frac54$, & $-$ & otherwise \\
		&				&			& $\wedge$ $\alpha>-\frac29$ & 	 &  \\ 		
$P_4$	&	$0$ 		&	$n$	& $-$ & $-$ & always \\ 	
$P_5$	&	$-1-m$	&	$n-1$	& $-$ & $-$ 
& $\Big[n>\frac{C_2}{4}$ \\
	&		&		&  &  
& $\wedge$ $-\frac19<\alpha<\frac13\Big]$ \\
	&		&		&  &  
& $\vee$ $\Big[1<n<\frac{C_1}{4}$   \\
	&		&		&  &  
& $\vee$ $n>\frac{C_2}{4}$ \\
	&		&		&  &  
& $\wedge$ $-\frac29<\alpha<-\frac19\Big]$ \\
	&		&		&  &  
& $\vee$ $\Big[1<n<\frac{C_2}{4}$ \\
	&		&		&  &  
& $\vee$ $n>\frac{C_1}{4}$ \\
	&		&		&  &  
& $\wedge$ $\alpha<-\frac29\Big]$ \\
$P_6$	&	$-1-m$		&	$n-1$	& $n<0$ $\vee$ $\frac12<n<1$ & $1<n<\frac54$ & otherwise \\
	&	&	& $\wedge$ $\alpha < \frac13$ & $\wedge$ $\alpha>-\frac29$ &	\\	
$P_7$	&	$-1-m$	&	$n-1$	& $-$ & $-$ & always \\ 	
$P_8$	&	$-1-m$	&	$n-1$	& $-$ & $-$ & $0<n<\frac{4-\sqrt{6}}{5}$ \\ 	&	&	&  &  & $\wedge$ $-\frac19<\alpha<\frac13$ \\ 	 
\hline \hline
\end{tabular}
\caption{The table shows the $m$ and $r$ parameter for all fixed points and the stability profiles with its possible values of $n$ for $f(R)=R- \gamma R^n$ model where $C_1 = (3-9\alpha)$ and $C_2 =\Big(3+ \frac{1}{4-9 \alpha } +\sqrt{\frac{9 \alpha  (9 \alpha -22)+73}{(4-9 \alpha )^2}}\,\Big)$. As shown in the table, the scaling solution point $P_5$ is saddle point when the range of $n$ does not lie on $0<n<1$ for all possible $\alpha$. This means that the $P_5$ is not compatible with the cosmologically viable of $f(R)$ dark energy of the model A when the interacting dark energy term, $Q=3\alpha H\rho_m$ is taken into account. }
\label{table:modelA}
\end{center}
\end{table}
%%%%%%%%%%%%%%%%%%%%%%%%%%%%%%%%%%%%%%%%%%%%%%%%%%%%%%%%%%%%%%%%%%%%%%%%%%%%%%%%%%%%%%%%%%%%%%%%%%%%%%%%%%%%%%%%%%%%%%%%%%%%%%%%%%%%%%%%%%%%%%%%%%%%%%%%%%
\begin{table}
\begin{center}
\begin{tabular}{ |c|c|c|c|c|c| }\hline
\multirow{2}{*}{}&	\multirow{2}{*}{$r=x_3/x_2$}	&\multirow{2}{*}{$m(r)$}	&	\multicolumn{3}{ |c| }{Stability} \\
\cline{4-6}
&	&	& Stable & Unstable & Saddle \\ \hline
$P_1$	&	$-2$		&	$\displaystyle{\frac{c+bc-2}{c}}$	& 	$c> \frac{25}{9}$	&  $-$ & otherwise \\
$P_2$	&	$-1-m$	&	$-1+bc$	& $-$ & $bc\rightarrow1^-$ & otherwise \\
		&			&				& 	 & $\alpha>-\frac29$ &  \\ 			
$P_3$	&	$-1-m$	&	$-1+bc$	& $-$ & $bc\rightarrow1^-$ & otherwise \\
		&			&				& 	 & $\alpha>-\frac29$ &  \\
$P_4$	&	$0$ 	&	$\displaystyle{b-1}$	& $0<b<1$ & $-$ & otherwise \\
$P_5$	&	$-1-m$	&	$-1+bc$	& $-$ & $-$ & $\Big[bc >\frac{C_2}{4}$ \\  	&	&	&   &   & $\wedge$ $-\frac19<\alpha<\frac13\Big]$ \\ 
	&	&	&   &   & $\vee$ $\Big[1<bc<\frac{C_1}{4}$ \\ 
	&	&	&   &   & $\wedge$ $-\frac29<\alpha<-\frac19\Big]$ \\ 
	&	&	&   &   & $\vee$ $\Big[bc\to 1^\pm$ \\ 
	&	&	&   &   & $\wedge$ $\alpha =-\frac29\big]$ \\ 
$P_6$	&	$-1-m$	&	$-1+bc$	& $bc\rightarrow1^-$ & $bc\rightarrow1^+$ & otherwise \\  	
$P_7$	&	$-1-m$	&	$-1+bc$	& $-$ & $-$ & always \\  	  	
$P_8$	&	$-1-m$	&	$-1+bc$	& $-$ & $-$ & $0<bc<\frac{4-\sqrt{6}}{5}$ \\ 	&	&	&  &  & $\wedge$ $c\geq 1$ \\  	  	 
\hline \hline
\end{tabular}
\caption{The table shows the $m$ and $r$ parameter for all fixed points and the stability profiles with its possible values of $b$ and $c$ from the model $f(R)=(R^b-\Lambda)^c$ where $C_1 = (3-9\alpha)$ and $C_2 =\Big(3+ \frac{1}{4-9 \alpha } +\sqrt{\frac{9 \alpha  (9 \alpha -22)+73}{(4-9 \alpha )^2}}\,\Big)$.}
\label{table-modelB}
\end{center}
\end{table}
%%%%%%%%%%%%%%%%%%%%%%%%%%%%%%%%%%%%%%%%%%%%%%%%%%%%%%%%%%%%%%%%%%%%%%%%%%
%%%%%%%%%%%%%%%%%%%%%%%%%%%%%%%%%%%%%%%%%%%%%%%%%%%%%%%%%%%%%%%%%%%%%%%%%%

%%%%%%%%%%%%%%%%%%%%%%%%%%%%%%%%%%%%%%%%%%%%%%%%%%%%%%%%%%%%%%%%%%%%%%%%%%%%%%%%%%%%%%%%%%%%%%%%%%%%%%%%%%%%%%%%%%%%%%%%%%%%%%%%%%%%%%%%%%%%%%%%%%%%
\subsection{Model B: $f(R) = \big( R^b - \Lambda\big)^c$ with $c \geq 1$ and  $bc\approx 1$}
The model B is proposed by Ref. \cite{Amendola:2006we,Amendola:2007nt} to generalize the LCDM and recover the local gravity constraints in GR. In this model, the $m$ and $r$ parameters are given by
\begin{align}
m&=\frac{(b c-1) R^b-b \Lambda +\Lambda }{R^b-\Lambda }\\
r&=-\frac{b c R^b}{R^b-\Lambda }.
\end{align}
Then the $m$ parameter can be written as a function of $r$ via,
\begin{align}
m(r)=\left(\frac{1-c}{c}\right)r+b-1.
\end{align}
We note that under limit $c\ge1$ and $bc\approx1$ the model B will be reduced to the LCDM with $m=0$ and provide viable cosmological evolution \cite{Amendola:2006we,Amendola:2007nt}. For the fixed points $P_{2,3,5,6,7,8}$, we found $r=-1-m$\,. The value of $m$ in this model is written by,
\begin{align}
m=-1+bc\,.
\end{align}
The explicit forms of all fixed points and their eigenvalues in model B are give by
\allowdisplaybreaks
\begin{eqnarray}
&\bullet& P_1 = (0,-1,2,0)\,,
\nonumber\\
&& \Omega_\text{m}=0,\qquad w_\text{eff}=-1.
\nonumber\\
&&{\rm eigenvalues}~:~\frac{6\pm c \left(-3 b-3 + \sqrt{\left(b-\frac{2}{c}+1\right)\left(25 b-\frac{50}{c}+9\right)}\,\right)}{2 (b c+c-2)}\,,~3(3\alpha-1)\,,~0\,.
\\
&\bullet& P_2 = (-1-9\alpha,0,0,0)\,,
\nonumber\\
&&\Omega_\text{m} = 2+9\alpha,\qquad w_\text{eff}=\frac{1}{3}\,,
\nonumber\\
&&{\rm eigenvalues}~:~-3 (3 \alpha -1)\,,~-9 \alpha -2\,,~\frac{9 \alpha +4 b c-3}{b c-1}\,,~0\,,
\\
&\bullet& P_3 = (1,0,0,0).
\nonumber\\
&&\Omega_\text{m} = 0,\qquad w_\text{eff}=\frac{1}{3}
\nonumber\\
&&{\rm eigenvalues}\quad\,:~5\,,~\frac{4 b c-5}{b c-1}\,,~9 \alpha +2 \,.
\\
&\bullet& P_4 = (-4,5,0,0)\,,
\nonumber\\
&&\Omega_\text{m} = 0,\qquad w_\text{eff}=\frac{1}{3}
\nonumber\\
&&{\rm eigenvalues}\quad\,:~-5\,,~\frac{4 b }{b-1}\,,~3(3\alpha-1)\,,~0\,.
\\
&\bullet& P_5 = \left(\frac{(3-9 \alpha ) (b c-1)}{b c},\frac{-9 \alpha -4 b c+3}{2 b^2 c^2},\frac{9 \alpha +4 b c-3}{2 b c},0\right)
\nonumber\\
&& \Omega_\text{m} = \frac{9 \alpha +b c (-27 \alpha +2 (9 \alpha -4) b c+13)-3}{2 b^2 c^2},\qquad w_\text{eff}=\frac{1 -3 \alpha -b c}{b c},
\nonumber\\
&&{\rm eigenvalues}\quad\,:~\frac{1}{4 b c (b c-1)}\Big( -9 \alpha -3 b c (\alpha  (6 b c-9)+1)+3 
\nonumber\\
&&\qquad\qquad\qquad\qquad ~\pm \sqrt{\big(b c-1\big)\big(4 (8-9 \alpha )^2 b^3 c^3-4 p_B b^2 c^2+3 q_B b c -81 (1-3 \alpha )^2 \big)}\Big)\,,
\nonumber\\
&&\qquad\qquad\qquad\qquad   ~3(1-3 \alpha)\, ,~0\,,
\nonumber\\
&&\qquad{\rm where}~\, p_B = (9 \alpha  (54 \alpha -55)+152)\,,~q_B = (3 \alpha -1) (261 \alpha -139)\,,
\\
&\bullet& P_6 = \left(\frac{4-2 b c}{2 b c-1}\,,\,\frac{5-4 b c}{2 b^2 c^2-3 b c+1}\,,\,\frac{b c (4 b c-5)}{(b c-1) (2 b c-1)}\,,\, 0\right),
\nonumber\\
&&\Omega_\text{m} = 0,\qquad w_\text{eff}=\frac{2}{1-2 b c}+\frac{2}{3 (b c-1)}-1
\nonumber\\
&&{\rm eigenvalues}\quad\,:~\frac{5-4 b c}{b c-1}\,,~-\frac{2 b c (b c-2)}{(b c-1) (2 b c-1)}\,,
\nonumber\\
&&\qquad\qquad\qquad\qquad \frac{9 \alpha +2 (9 \alpha -4) b^2 c^2+(13-27 \alpha ) b c-3}{(b c-1) (2 b c-1)}\,,~0\,,
\\
&\bullet& P_7 = (0,0,0,1)
\nonumber\\
&&\Omega_\text{m} = 0,\qquad w_\text{eff}=\frac{1}{3}
\nonumber\\
&&{\rm eigenvalues}\quad\,:~4\,,~4\,,~-1\,,~1+ 9 \alpha\,.
\\
&\bullet& P_8 = \left(4-\frac{4}{b c}\,,\,\frac{2-2 b c}{b^2 c^2}\,,\,2-\frac{2}{b c}\,,\,-\frac{2}{b^2 c^2}+\frac{8}{b c}-5\right)
\nonumber\\
&& \Omega_\text{m} = 0,\qquad w_\text{eff}=-1 +\frac{4}{3bc}
\nonumber\\
&&{\rm eigenvalues}\quad\,:~4\,,~\frac{b^2 c^2-2 b c \pm b c \sqrt{3\big(27 b^2 c^2-44 b c+12\big)}}{2 b^2 c^2}\,,~1+9 \alpha\,.
\end{eqnarray}
For the parameters $m$ and $r$ in the model B are shown in table \ref{table-modelB}. To study the stability of the fixed points in the model B, we consider the limits of $bc\to 1$ via $bc\to 1^\pm$\,. As shown in table \ref{table-modelB}, we find that there are two fixed points that can represent the late time accelerating universe, i.e., $P_1$ and $P_6$. The de Sitter fixed point $P_1$ is still compatible with the conditions $c\geq 1$ and $bc\approx 1$ for model B while the curvature fluid dominated point $P_6$ is stronger compatible to the conditions of the model B than the point $P_1$. In addition, the scaling solution $P_5$ also compatible with $c\geq 1$ and $bc\approx 1$ conditions for all possible values $\alpha < 1/3$. We can see the ranges from table \ref{table-modelB} where $C_1\approx 0.32,\,3.99,\,4.89$ and $C_2\approx 7.67,\,5.16,\,5.00$ for $\alpha\approx 1/3,\,-1/9,\,-2/9$ respectively. In particular, we found that the condition $bc\approx 1$ is more valid when the $\alpha$ becomes more negative. The interacting dark energy does not change cosmologically viable profile of the model B. 
%%%%%%%%%%%%%%%%%%%%%%%%%%%%%%%%%%%%%%%%%%%%%%%%%%%%%%%%%%%%%%%%%%%%%%%%%%%%%%%%%%%%%%%%%%%%%%%%%%%%%%%%%%%%%%%%%%%%%%%%%%%%%%%%%%%%%%%%%%%%%%%%%%%%
\begin{table}
\begin{center}
\begin{tabular}{ |c|c|c|c|c|c| }\hline
\multirow{2}{*}{}&	\multirow{2}{*}{$r=x_3/x_2$}	&\multirow{2}{*}{$m(r)$}	&	\multicolumn{3}{ |c| }{Stability} \\
\cline{4-6}
&	&	& Stable & Unstable & Saddle \\ \hline
$P_1$	&	$-2$		&	$\mu$	& 	$\frac{16}{25}\leq \mu<1$	&  $-$ & $\mu>1$  \\
		&			&				& $\wedge$ $\alpha < \frac13$	   &  & 
$\wedge$ $\alpha < \frac13$ \\
$P_2$	&	$-1-m$	&	$\mu^{-\frac{1}{2n}}$	& $-$ & $\mu<[\frac{4}{(9\alpha+1)}]^{2n}$ & otherwise \\
		&			&				& 	 & $\wedge$ $\alpha<-\frac29$ &  \\   	
$P_3$	&	$-1-m$	&	$\mu^{-\frac{1}{2n}}$	& $-$ & $\mu<4^{2n}$ & otherwise \\
		&			&				& 	 & $\wedge$ $\alpha>-\frac29$ &  \\  
$P_4$	&	$0$ 	&	$-\mu$	& $-$ & $-$ & always \\ 
$P_5$	&	$-1-m$	&	$\mu^{-\frac{1}{2n}}$	& $-$ & $-$ & $\Big[0<\mu < \big[\frac{C_3}{4}\big]^{2 n}$  \\ 
	&	&	&   &   & $\wedge$ $-\frac19<\alpha<\frac13\Big]$ \\ 
	&	&	&   &   & $\vee$ $\Big[\mu>\big[\frac{4}{(9\alpha+1)}]^{2n}$ \\ 
	&	&	&   &   & $\wedge$ $-\frac29<\alpha<-\frac19\Big]$ \\ 
$P_6$	&	$-1-m$	&	$\mu^{-\frac{1}{2n}}$	& $n>0$ $\wedge$ $0<\mu<1$ & $\mu>4^{2n}$ & otherwise \\
		&			&				& $\wedge$ $\alpha<\frac13$	   &  &  \\  	 
%		&			&				& or $\mu < \left[(C -9 \alpha +3)/4\right]^{2 n}$	   &  &  \\ 
%				&			&				& and $1/3<\alpha<4/9$	   &  &  \\
$P_7$	&	$-1-m$	&	$\mu^{-\frac{1}{2n}}$	& $-$ & $-$ & always \\ 
$P_8$	&	$-1-m$	&	$\mu^{-\frac{1}{2n}}$	& $-$ & $-$ & $\mu>0$ $\wedge$ $n>0$ \\  	  	 
\hline \hline
\end{tabular}
\caption{Table shows $m$ and $r$ parameters and the possible of stability of all fixed points from the model C, $f(R)=R-\lambda R_c\left[1-\left(\frac{R_c}{R}\right)^{2n}\right]$ with $\mu\equiv \frac{2n(2n+1)}{\lambda^{2n}}$ and $C_3=-9\alpha + 3 + \sqrt{81 \alpha ^2-198 \alpha +73}$.}
\label{table-modelC}
\end{center}
\end{table}
%%%%%%%%%%%%%%%%%%%%%%%%%%%%%%%%%%%%%%%%%%%%%%%%%%%%%%%%%%%%%%%%%%%%%%%%%%%%%%%%%%%%%%%%%%%%%%%%%%%%%%%%%%%%%%%%%%%%%%%%%%%%%%%%%%%%%%%%%%%%%%%%%%%%
\subsection{Model C: $f(R) = R - \lambda\,R_c\,\Big[ 1 + \left(R_c/R\right)^{2n}\Big]$ with $n,\,\lambda >0$ and $R>R_c$}
We consider the model C: $f(R) = R - \lambda\,R_c\,\Big[ 1 + \left(R_c/R\right)^{2n}\Big]$ with $n,\,\lambda >0$ and $R>R_c$ in this section where the model C is an approximation of the Hu and Sawicki model, $f(R) = R-\lambda R_c\,(R/R_c)^{2n}/\big[ 1+ (R/R_c)^{2n}\big]$ \cite{Hu:2007nk} and the Starobinsky model $f(R) = R -\lambda R_c\big[ 1 - \big(1+ R^2/R_c^2 \big)^{-n}\big]$ \cite{Starobinsky:2007hu} in the $R>R_c$ limit. Those two models were proposed to reproduce the local gravity constraints of $f(R)$ gravity, while still provide the accelerating universe solution at the large scale. The parameters $m$ and $r$ and their related quantities in the dynamical system analysis in model C are given by
\begin{align}
m&= \frac{2n(2n+1)}{\lambda^{2n}}(-r-1)^{2n+1} 
\nonumber\\
&= \mu(-r-1)^{2n+1}\,,\quad \mu \equiv \frac{2n(2n+1)}{\lambda^{2n}}\,,
\\
r&= -1-\lambda\frac{R_c}{R}\,.
\end{align}
We can write the explicit forms of all fixed points in the model C as
\allowdisplaybreaks
\begin{eqnarray}
&\bullet& P_1 = (0,-1,2,0)\,,
\nonumber\\
&& \Omega_\text{m}=0,\qquad w_\text{eff}=-1.
\nonumber\\
&&{\rm eigenvalues}~:~-\frac{3}{2} \pm \sqrt{\frac{25 \mu-16}{4\mu}}\,,~3(3\alpha-1)\,,~0\,.
\\
&\bullet& P_2 = (-1-9\alpha,0,0,0)\,,
\nonumber\\
&&\Omega_\text{m} = 2+9\alpha,\qquad w_\text{eff}=\frac{1}{3}\,,
\nonumber\\
&&{\rm eigenvalues}~:~-3 (3 \alpha -1)\,,~-9 \alpha -2\,,~ 4 + (9 \alpha +1) \mu^{\frac{1}{2 n}}\,,~0\,,
\\
&\bullet& P_3 = (1,0,0,0).
\nonumber\\
&&\Omega_\text{m} = 0,\qquad w_\text{eff}=\frac{1}{3}
\nonumber\\
&&{\rm eigenvalues}\quad\,:~5\,,~4-\mu ^{\frac{1}{2 n}}\,,~9 \alpha +2\,,~0 \,.
\\
&\bullet& P_4 = (-4,5,0,0)\,,
\nonumber\\
&&\Omega_\text{m} = 0,\qquad w_\text{eff}=\frac{1}{3}
\nonumber\\
&&{\rm eigenvalues}\quad\,:~-5\,,~4-\frac{4}{\mu }\,,~3(3\alpha-1)\,,~0\,.
\\
&\bullet& P_5 = \left(\frac{3-9 \alpha }{\mu ^{\frac{1}{2 n}}+1}\,,\,-\frac{9 \alpha +4 \mu ^{-\frac{1}{2 n}}+1}{2 \left(\mu ^{-\frac{1}{2 n}}+1\right)^2}\,,\,\frac{1}{2} \left(9 \alpha +\frac{3-9 \alpha }{\mu ^{\frac{1}{2 n}}+1}+1\right)\,,\,0\right)
\nonumber\\
&& \Omega_\text{m} = \frac{18 \alpha +3 (3 \alpha -1) \mu ^{\frac{1}{2 n}}+2 \mu ^{1/n}-8}{2 \left(\mu ^{\frac{1}{2 n}}+1\right)^2}\,,\qquad w_\text{eff}=\frac{3 \alpha -1}{\mu ^{\frac{1}{2 n}}+1}-3 \alpha\,,
\nonumber\\
&&{\rm eigenvalues}\quad\,:~\frac{\mu ^{\frac{1}{2 n}}}{4 \left( 1 + \mu ^{\frac{1}{2 n}}\right)} \Bigg[-3-\alpha  \left(18 \mu ^{-\frac{1}{2 n}}+ 9\right) 
\nonumber\\
&&\qquad\qquad\qquad\qquad \pm \mu ^{-\frac{1}{2 n}} \sqrt{256 -81 \alpha ^2 p_C -18 \alpha q_C -31 \mu ^{1/n}+160 \mu ^{\frac{1}{2 n}}-16 \mu ^{\frac{3}{2 n}}}\,\Bigg]\,,
\nonumber\\
&&\qquad\qquad\qquad\qquad   ~3(1-3 \alpha)\, ,~0\,,
\nonumber\\
&&\quad{\rm where}~\, p_C = \left(7 \mu ^{1/n}+12 \mu ^{\frac{1}{2 n}}-4\right) \,,~q_C = \left(-11 \mu ^{1/n}-14 \mu ^{\frac{1}{2 n}}+8 \mu ^{\frac{3}{2 n}}+32\right)\,,
\\
&\bullet& P_6 = \left(\frac{2 \left(\mu ^{\frac{1}{2 n}}-1\right)}{\mu ^{\frac{1}{2 n}}+2}\,,\,\frac{\mu ^{1/n}-4 \mu ^{\frac{1}{2 n}}}{\mu ^{\frac{1}{2 n}}+2}\,,\,\frac{4 -\mu ^{1/n}+3 \mu ^{\frac{1}{2 n}}}{\mu ^{\frac{1}{2 n}}+2}\,,\, 0\right),
\nonumber\\
&&\Omega_\text{m} = 0,\qquad w_\text{eff}= -3 + \frac{2}{3} \mu ^{\frac{1}{2 n}}+\frac{4}{\mu ^{\frac{1}{2 n}}+2}
\nonumber\\
&&{\rm eigenvalues}\quad\,:~\mu ^{\frac{1}{2 n}}-4\,,~\frac{2 \left(\mu ^{1/n}-1\right)}{\mu ^{\frac{1}{2 n}}+2}\,,~\frac{9 \alpha  \left(\mu ^{\frac{1}{2 n}}+2\right)+2 \mu ^{1/n}-3 \mu ^{\frac{1}{2 n}}-8}{\mu ^{\frac{1}{2 n}}+2}\,,~0\,,
\\
&\bullet& P_7 = (0,0,0,1)
\nonumber\\
&&\Omega_\text{m} = 0,\qquad w_\text{eff}=\frac{1}{3}
\nonumber\\
&&{\rm eigenvalues}\quad\,:~4\,,~4\,,~-1\,,~1+ 9 \alpha\,.
\\
&\bullet& P_8 = \left(\frac{4}{\mu ^{\frac{1}{2 n}}+1}\,,\, -\frac{2 \mu ^{\frac{1}{2 n}}}{\left(\mu ^{\frac{1}{2 n}}+1\right)^2}\,,\, 
\frac{2}{\mu ^{\frac{1}{2 n}}+1}\,,\, \frac{\mu ^{1/n}-2 \mu ^{\frac{1}{2 n}}-5}{\left(\mu ^{\frac{1}{2 n}}+1\right)^2}\right)
\nonumber\\
&& \Omega_\text{m} = 0,\qquad w_\text{eff}= \frac{1}{3}-\frac{4}{3 \left(\mu ^{\frac{1}{2 n}}+1\right)}\,,
\nonumber\\
&&{\rm eigenvalues}\quad\,:~4\,,~\frac{1\pm\mu ^{1/n} \left(1+ \sqrt{3\big(27 \mu ^{-2/n}+42 \mu ^{-1/n}+64 \mu ^{-\frac{3}{2 n}}-5\big)}\right)}{2 \left(1+ \mu^{\frac{1}{2 n}}\right)^2}\,,
\nonumber\\
&&\qquad\qquad\qquad\qquad   ~1+9 \alpha\,.
\end{eqnarray}
The values of the $m$ and $r$ parameters for all fixed points in the model C as well as their stability are summarized in table \ref{table-modelC}.

We close this section by discussing the influence of the interacting dark energy in the $f(R)$ gravity model C. According to our results given in table \ref{table-modelC}, we discover that the cosmological viability of the $f(R) = R - \lambda\,R_c\,\big[ 1 + \left(R_c/R\right)^{2n}\big]$ model is still compatible with the standard LCDM when the interacting dark energy is taken into account. However, we find some inconsistent predictions of the $\lambda$ parameter of the model C between the non-interacting and the interacting  dark energy at the de Sitter fixed point $P_1$. For non-interacting dark energy in $f(R)$ gravity, it has been shown in the Ref. \cite{Tsujikawa:2007xu} that the $\lambda$ parameter of the model C is constrained by at $\lambda > 8\sqrt{3}/9$ for $n=1$ whereas the interacting dark energy provides the constraint $16/25<\mu<1$ for the stable node of the de Sitter point $P_1$. This leads to $\sqrt{6} < \lambda < \sqrt{75/8}$ at $n=1$ for $\mu \equiv 2n(2n+1)/\lambda^{2n}$. As the result, we find that the interacting dark energy increase the lower bound and provide the upper limit of the $\lambda$ parameter in the model C.
%%%%%%%%%%%%%%%%%%%%%%%%%%%%%%%%%%%%%%%%%%%%%%%%%%%%%%%%%%%%%%%%%%%%%%%%%%
%%%%%%%%%%%%%%%%%%%%%%%%
\section{Conclusion}
In this work, we have qualitatively studied the cosmological dynamics of the interacting dark energy and dark matter in the viable models of $f(R)$ gravity by using the standard dynamical system approach. For simplicity, the simple interacting dark energy and dark matter model, $Q=3\alpha H\rho_m$ is used in the present study. From our set up of an autonomous system of cosmological dynamical equations, we have obtained total eight fixed points and there are six and two fixed points for absent and apparent of the radiation energy density, respectively as found in the literature for non-interacting dark energy case with additional parameter $\alpha$ in the fixed point solutions. In the present work, the cosmological viability of the $f(R)$ gravity is found with an additional constraints from the interacting dark energy parameter $\alpha <1/3$.

Next, three models named A, B and C of the viable $f(R)$ gravity are investigated in detail for interacting dark energy system where all those three models in this work have been systematically constructed with possible ranges of its model parameters from the dynamical system analysis for the interacting dark energy framework. It is worth seeing how the interacting affect to the cosmological viability of the $f(R)$ gravity models. 

As the result, we found that the model A, $f(R)= R -\gamma R^n$ with $\gamma >0$ and $0<n<1$ is not cosmologically viable in the presence of the interacting dark energy. The stability condition of the de Sitter fixed point $P_1$ requires $32/25<n<2$ and this is contradict to the viable condition of the model A. Interestingly, the scaling solution fixed point $P_5$ yields $n>1.26,\,1.92$ for $\alpha\approx -1/9,\,1/3$, respectively. For matter dominated epoch, we can see that the possible range of the $n$ lies outside the $0<n<1$ condition for cosmological viability of the $P_5$ point in the model A, when it becomes saddle point, see table \ref{table:modelA}; while, the model B, $f(R)=\big( R^b - \Lambda\big)^c$ is still cosmologically viable for $f(R)$ dark energy model. All stability of the fixed points of the model B in the presence of the interacting dark energy are compatible with the constraints $c\geq 1$ and $bc\approx 1$\,. Considering the model C with $f(R) = R - \lambda\,R_c\,\big[ 1 + \left(R_c/R\right)^{2n}\big]$, the result shows that this model is also cosmologically viable in the presence of the interacting dark energy. However, we found some inconsistent predictions of a coupling $\lambda$ between the non-interacting and the interacting one. At $n=1$, the non-interacting dark energy reveals $\lambda >8\sqrt{3}/9$ whereas our result shows the constraint $\sqrt{6}<\lambda<\sqrt{75/8}$.    

Although there are many viable $f(R)$ models explaining the dark energy problem in cosmology, our qualitative results found in the present work could be guidelines for more detail study and are used as complementary constraints on the viable $f(R)$ models in addition to the other cosmological constraints on $f(R)$ theories. All these $f(R)$ gravity models are significantly distinguishable by the evolution of the cosmological perturbations. In particular, the study of the formation of structure in the Universe is sensitive to the interacting dark energy and dark matter in the $f(R)$ gravity. More importantly, the presence of the interacting $f(R)$ dark energy and dark matter in the early phase of the universe might affect the epoch of matter-radiation equality. This pattern of anisotropies could be calculated in terms of the growth of the structure formation in the interacting $f(R)$ gravity framework. Remarkably, the interaction between $f(R)$ dark energy and dark matter at the early stage of the cosmological evolution might describe the discrepancy of the Hubble tension parameter between local measurement and another one from the cosmic microwave background result. Moreover, the matter perturbation and the local gravity approximations of the interacting dark energy in $f(R)$ gravity is worth for further study in order to explain the inconsistent predictions between the non-interacting and the interacting $f(R)$ dark energy models. By the way, other models of interacting dark energy in $f(R)$ gravity might be interesting for further study to extract more physical implications and cosmological consequences.
%%%%%%%%%%%%%%%%%%%%%%%%

\acknowledgments
D. Samart is financially supported by National Astronomical Research Institute of Thailand (NARIT). B. Silasan is supported by the Development and Promotion of Science and Technology Talents Project (DPST). P. Channuie acknowledged the Mid-Career Research Grant 2020 from National Research Council of Thailand under a contract No. NFS6400117 and is financially supported by the new strategic research project (P2P), Walailak University, Thailand.

\end{document}